\documentclass[prl,twocolumn,showpacs]{revtex4}
\usepackage{amsmath,amssymb}
\usepackage{epsfig}

\begin{document}

\title{Monte Carlo Test of the Classical Theory for Heterogeneous Nucleation Barriers}

\author{David Winter, Peter Virnau, K. Binder}
\affiliation{Institut f\"{u}r Physik, Johannes
Gutenberg-Universit\"{a}t,\\
 D-55099 Mainz, Staudinger Weg 7, Germany}

\date{\today}
\begin{abstract}
Flat walls facilitate the condensation of a supersaturated vapor:
Classical theory of heterogeneous nucleation predicts that the free energy barrier $\Delta
F_{\rm het}^*$ which needs to be overcome for the formation of sphere-cap
shaped nucleation seeds is smaller than the barrier $\Delta F^*_{\rm hom}$
for spherical droplets in the bulk by a factor $0<f(\theta)<1$,
which only depends on the contact angle $\theta$.
In this letter we compute both $\Delta F^*_{\rm hom}$ 
and $\Delta F^*_{\rm het}$ from Monte Carlo simulations and test the theory
for the lattice gas model (for which $\theta$ can be readily controlled).
Even though the theory is only based on macroscopic arguments, 
it is shown to hold for experimentally relevant nanoscopic nucleation seeds
($20\leq\Delta F^*_{\rm hom}/k_BT\leq 200)$ if (independently
estimated) line tension effects are considered.
\end{abstract}


\maketitle

Nucleation \cite{1,2,3,4,5} is a ubiquitous process which governs such diverse
phenomena as the formation of rain drops or snow flakes in the atmosphere,
the crystallization of proteins and even the formation of industrially
relevant polymer foams. Even though considerable empirical knowledge has been 
gathered, e.g., on which materials are suited as seed particles for making clouds rain \cite{9}
or as nucleation agents to start precipitation in metallurgy \cite{10}, many questions
remain.
Reasons for this partial shortcoming are manifold: Experiments are
very difficult to control. Homogeneous nucleation in the bulk needs to be separated
from heterogeneous nucleation at walls or impurities, and small changes in parameters
typically lead to large changes in corresponding nucleation rates. Nucleation
sites often only consist of a few hundred or thousand atoms and can only be observed indirectly.
Unfortunately, theoretical progress is also hampered by these difficulties.
Simulations on the
other hand offer full control and are at least in principle able to bridge the gap between
theory and experiment and, indeed, significant progress has been achieved in recent years \cite{12,27,29}.

According to Classical Nucleation Theory \cite{1,2}, the formation of a (spherical) nucleation seed in the bulk is understood
in terms of two competing factors: a volume term which seeks to expand the seed and an opposing surface term $F_{s} = 4 \pi R^2 \gamma$:
\begin{equation}
\Delta F(R)=-\Delta \mu ( \rho_{l} - \rho_{v} ) \frac{4\pi R^3}{3} + 4 \pi R^{2} \gamma.
\label{CNT1}
\end{equation}
$R$ is the radius of the droplet (or bubble), $\gamma$ the interfacial tension (of a macroscopically
flat interface), $\Delta \mu= \mu-\mu_{coex}$ the difference in chemical potential relative to the coexistence value,
and $\rho_{l,v}$ the densities of the coexisting phases. The free energy barrier $\Delta F^{*}$ can now be obtained
by determining the maximum of Eq.~\eqref{CNT1} at $R^{*}$. Interestingly, at $R^{*}$ we also obtain
\begin{equation}
\Delta F^{*}=\frac{1}{3}F_{s}(R^{*}).
\label{CNT2}
\end{equation}
Hence, the nucleation barrier can directly be inferred from the knowledge of the surface free energy of a droplet 
or bubble. Note that the
spontaneous formation of a critical nucleus of the new phase in
the bulk requires a free energy
barrier $\Delta F^*_{\rm hom}$ of the order $20 \leq \Delta
F^*_{\rm hom}/k_BT \leq 200$. Compared to thermal fluctuations
(that one can easily detect e.g. by
scattering experiments) this is a rare event.

\begin{figure}[t!bp]
\includegraphics[width=0.8\linewidth]{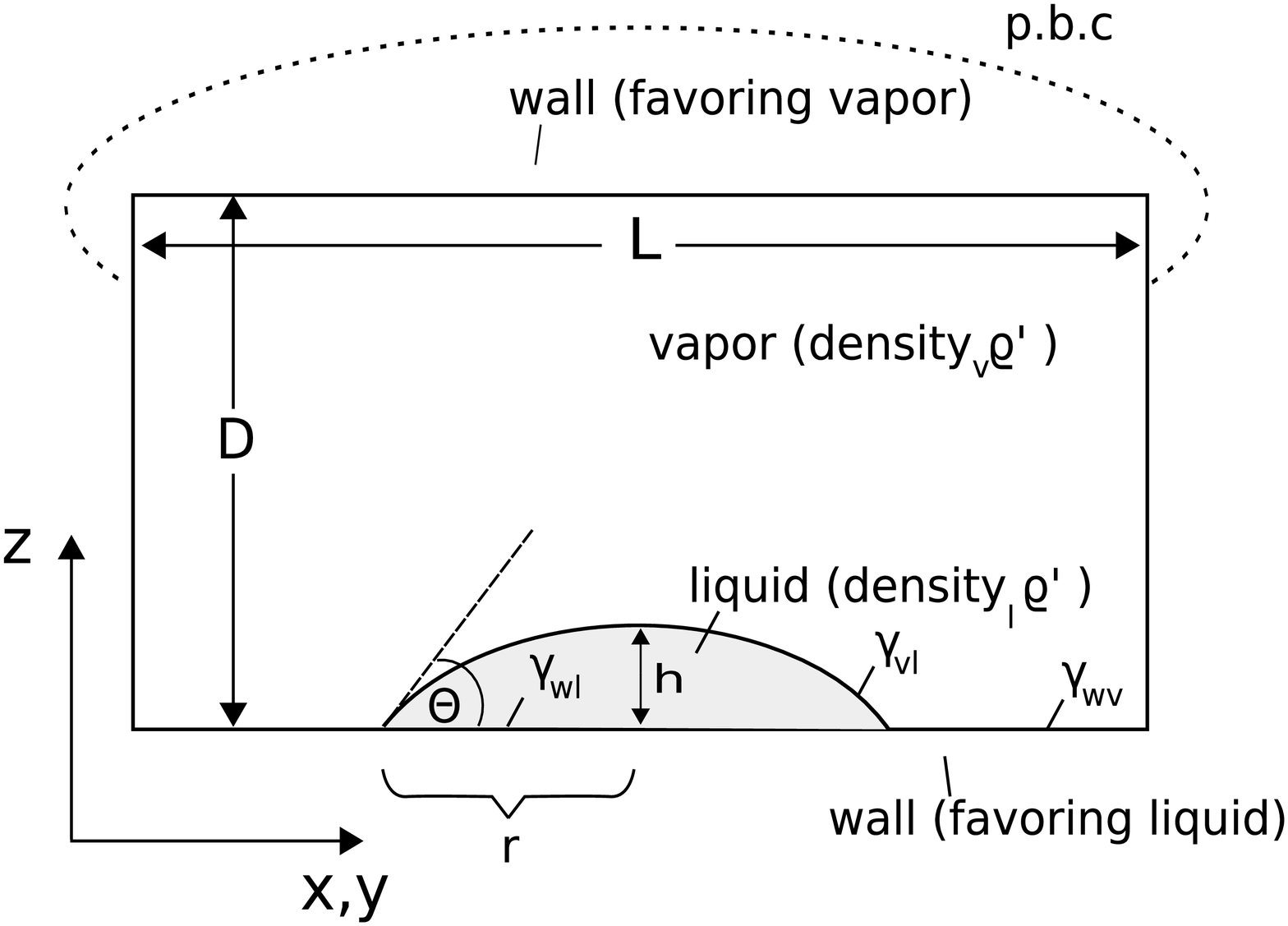}
\includegraphics[width=0.5\linewidth]{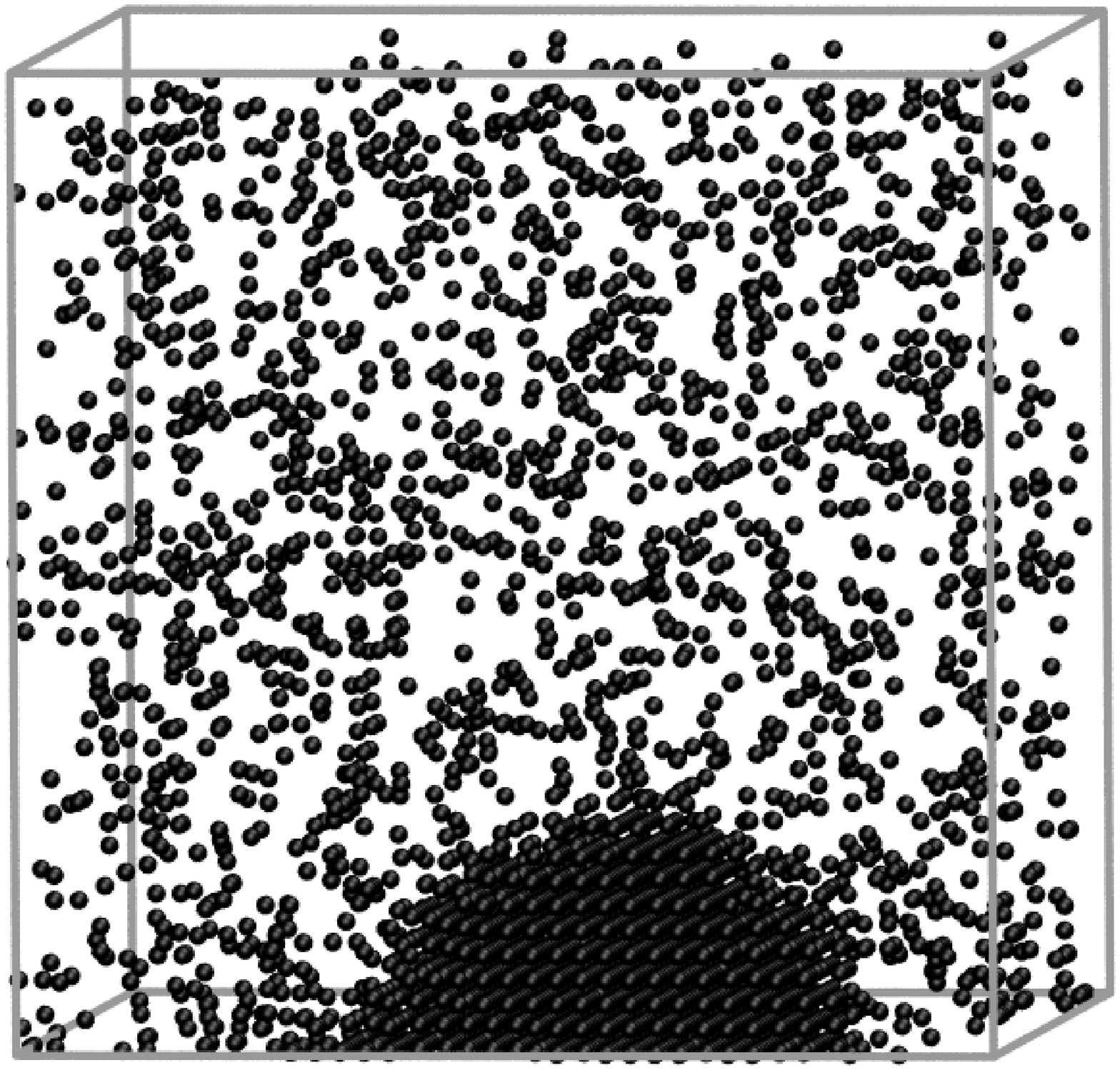}
\caption{ a) Schematic drawing of the system which we use to study stable
wall-attached droplets in thermal equilibrium. The sphere-cap
shaped droplet has height $h$ and covers a circle of radius $r$ at
the wall, with $r=R \sin \theta$, $h=R(1-\cos\theta)$. $R$ denotes
the radius of curvature and $\theta$ the contact angle. For
$R<\infty$, the density $\rho'_v$ of the vapor which coexists with the
liquid droplet (at the same chemical potential $\mu > \mu_{\rm coex})$
exceeds $\rho_v$ (Gibbs-Thomson effect \cite{6,7}). b) Typical
snapshot picture of a lattice gas system at $k_BT/J=3.0$,
$\rho=0.065$, $L=D=40$ (measured in units of the lattice
spacing), and $H_1/J=0$ $(\theta=90^o)$. Occupied lattice sites are
highlighted by dots.} \label{fig1}
\end{figure}

The presence of a wall facilitates nucleation and reduces the barrier.
For macroscopic droplets (see Fig.~\ref{fig1}), the contact angle $\theta$ between the droplet and the wall
is given by the competition of the vapor-liquid
($\gamma_{v \ell})$, wall-vapor ($\gamma_{wv})$ and wall-liquid
($\gamma_{w \ell})$ interface free energies \cite{13}
\begin{equation}\label{Young}
\gamma_{v \ell} \cos \theta=\gamma_{wv} - \gamma_{w \ell},
\end{equation}
for incomplete wetting conditions ($\gamma_{v \ell} > \gamma_{w v}
- \gamma_{w \ell})$ \cite{14}. Eq.~\eqref{Young} is known as Young's equation and
can be derived by considering the forces acting at the three phase boundary.
For small droplets, a
correction due to the line tension ($\tau$) of the contact line arises
\cite{7} (which has the length $2 \pi r$, where $r=R\sin \theta$).

Considering Eq.~\eqref{Young}
and the geometry of the sphere-cap (see caption of Fig.~\ref{fig1}) we obtain Turnbull's estimate for the surface free energy \cite{6,7,8}:
\begin{eqnarray}
&& F_{s,het}=F_{s,hom} \cdot f(\theta), \nonumber\\ 
&&f(\theta)=(1-\cos\theta)^2(2+\cos\theta)/4.
\label{Turnbull1}
\end{eqnarray}
Intriguingly, $f(\theta)=V_{sphere-cap}/V_{droplet}$. Therefore, Eq.~\eqref{CNT2}
also holds for the heterogeneous case and we get:
\begin{equation}
\Delta F_{het}=\Delta F_{hom} \cdot f(\theta).
\label{Turnbull2}
\end{equation}
The nucleation barrier in the heterogeneous case is simply reduced by a factor of
$f(\theta)$. Note that this derivation is solely based on macroscopic considerations
and has to our knowledge never actually been tested. In this letter we will demonstrate with
Monte Carlo simulations of an Ising lattice gas model that this formula can
be applied to nanoscopic systems if line tension effects are considered. 
This provides a stringent test for both the classical theory of homogeneous and
heterogeneous nucleation.
For this purpose we propose a new method \cite{11,21} to measure surface
free energies (and hence nucleation barriers) of liquid droplets 
in the bulk and of droplets attached to walls over the experimentally 
relevant range. No ``atomistic'' identification of
which particles belong to the droplet and which to the vapor is
required. No bias potential to stabilize droplets of a particular
size \cite{12} is needed, and fluctuations of the droplet are not
constrained. We independently determine the droplet volume and the
chemical potential that characterizes the ({\it stable})
equilibrium between the small droplet and surrounding vapor in the
{\it finite system} and derive an expression for the line tension. 

We test Turnbull's formula \eqref{Turnbull2} with Monte Carlo simulations of the
standard nearest neighbor (Ising) lattice gas model 
on the simple cubic lattice: $\mathcal H=-J \sum S_i S_j, S_{i,j} = \pm 1$.
Phase coexistence between
saturated vapor at density $\rho_v$ and liquid $\rho_\ell$ occurs
at a (known) chemical potential $\mu_{\rm coex}$.
(In magnetic notation, the magnetic field $H=(\mu-\mu_{\rm
coex})/2$, and $\rho_\ell$, $\rho_v$ are related to the
spontaneous magnetization $m_{\rm coex}$ as $\rho_v=(1-m_{\rm
coex})/2$, $\rho_\ell =(1+m_{\rm coex})/2$.) 
Bulk systems are studied in cubic simulation boxes
with periodic boundary conditions in all directions.
For the heterogeneous case we choose a $L \times
L \times D$ geometry with periodic boundary conditions in $x$ and
$y$ direction. Surface fields $H_1$ and $H_D=-H_1$ act on
(the first layer of) the two free $L \times L$ surfaces. 
The temperature is set to $k_BT/J=3.0$ to stay away from the bulk critical temperature (at $k_BT/J\approx4.51$)
and above the roughening transition temperature \cite{18} (at $k_BT/J\approx2.45$).
Hence, the correlation length $\xi$ is less than a lattice
spacing and the interface free
energy does (to a good approximation) not depend on the interface orientation.


\begin{figure}[tbp]
\includegraphics[width=0.75\linewidth]{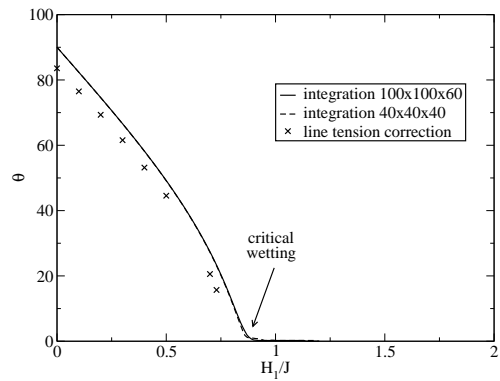}
\caption{Contact angle $\theta$ as derived from
Eq.~\eqref{eq3} plotted vs. $H_1/J$ at $k_BT/J=3.0$
for two lattice sizes (full and broken curves). Symbols show
the prediction if the estimates for
$\tau$ from our analysis of the droplet excess free energies and $r=5$ 
are considered (see below).} \label{fig2}
\end{figure}

Wetting for the Ising model has been thoroughly studied before \cite{15}
and by varying the surface field $H_1$ we can control the contact angle 
(Fig.~\ref{fig2}). Note that the local surface layer
magnetization $m_1=-(\partial f_s (T,H,H_1)/\partial H_1)_{T,H}$
\cite{16} ($f_s$ is the surface excess free energy per spin).
With $f_s (T,H \rightarrow 0^-, H_1)=
\gamma_{w v}$, $f_s(T, H \rightarrow 0^+, H_1)=\gamma_{w \ell}$ and Young's equation \eqref{Young},
we obtain $\theta$ from thermodynamic integration

\begin{equation} \label{eq3}
\cos \theta = (1/\gamma_{v \ell}) \int\limits_0^{H_1} (m_D-m_1)d
H'_1 \,\, .
\end{equation}
Here, we exploit the Ising symmetry $f_s(T,H \rightarrow 0^-, H_1)=f_s(T,H
\rightarrow 0^+,-H_1)$ to use the local
magnetization $m_D$ at the other surface at which $H_D=-H_1$ acts.
$\gamma_{v \ell}=0.434$ was given by the very accurate estimates of
Hasenbusch and Pinn \cite{17}.

\begin{figure}[tbp]
\includegraphics[width=0.9\linewidth]{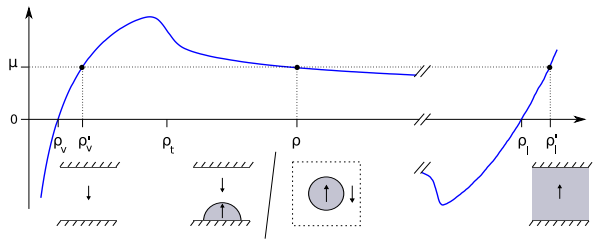}\vspace{0.2cm}
\caption{Chemical potential as a function of density. The density of the sphere-cap shaped droplet
at overall density $\rho$ is given by $\rho_l'$, the density of the environment by $\rho_v'$.
$\Delta F_s$ is obtained by integrating
the chemical potential over the density from $\rho_v'$ to $\rho$.
}
\label{fig_chem_pot}
\end{figure}

In the following we describe how to obtain the surface free energy as a function of the droplet radius.
First, we measure the chemical
potential (in the canonical ensemble) as a function of density (Fig.~\ref{fig_chem_pot}) by applying
an adaptation \cite{11,21} of the Widom test particle insertion method \cite{19}.
In the flat region of the isotherm depicted in Fig.~\ref{fig_chem_pot}, a single droplet is present
in the simulation box, which will be used for the determination of the surface free energy.
At small densities the
ascending branch of $\mu$ represents
pure vapor (with no single large droplet being present), and the subsequent rapid decrease
signifies the system size dependent transition towards a single droplet phase
\cite{22,23,24,25}.
At larger densities, the droplet region is delimited by a transition towards a
cylindrical droplet which is stabilized by the boundary conditions.
Note that for each system size, only a certain range of droplet sizes can be stabilized
and we have to simulate a whole range of box sizes to cover all radii.

As indicated in Fig.~\ref{fig_chem_pot} for each droplet state at density $\rho$, there exists 
one vapor state at $\rho_v'$ and one liquid state at $\rho_l'$ which have the same
chemical potential. Equilibrium conditions require that the chemical potential is constant
throughout the box. Therefore, the droplet has density $\rho_l'$ and is
surrounded by an environment with density $\rho_v'$. 
The surface free energy
of the droplet is simply given by the difference in free energy between the droplet and the vapor state
and can be determined by thermodynamic integration:

\begin{equation}
F_s(\rho)=\int_{\rho_v'}^\rho \mu(\rho') d\rho'.
\label{thermo_int}
\end{equation}
The number of particles in the box is conserved:

\begin{equation} \label{eq7}
\rho_v'(V-V_{drop})+\rho_l'V_{drop}=\rho V.
\end{equation}
\noindent
If we further require that the (average) droplet
is spherical in the bulk and sphere-cap shaped for the heterogeneous case, $V_{drop}$
becomes $4/3\pi R^3$ or $4/3\pi R^3 f(\theta)$, respectively, and
we obtain $R(\rho)$ and finally $F_s(R)$ and $F_s(R,\theta)$.

\begin{figure}[tbp]
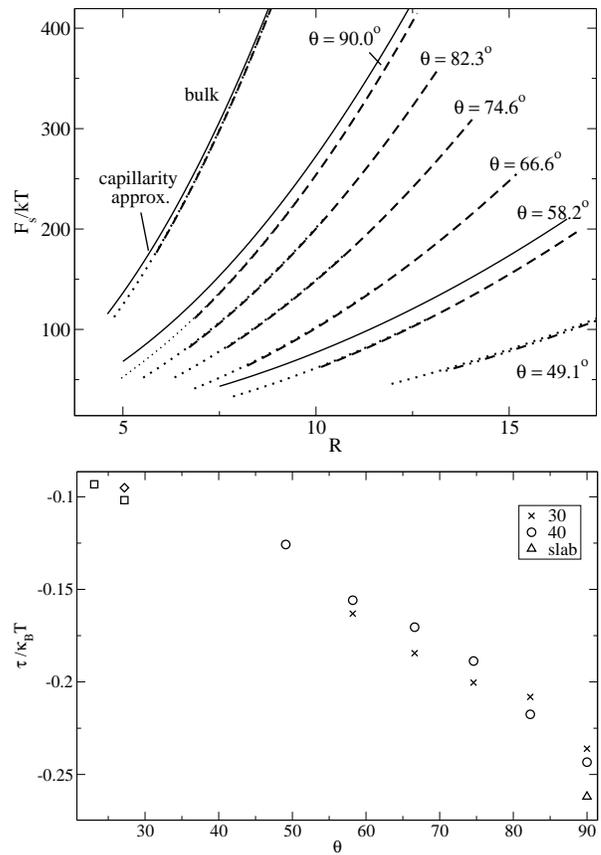

\includegraphics[width=0.9\linewidth]{uebersicht_freie_energien2.eps}\vspace{0.2cm}
\includegraphics[width=0.9\linewidth]{lt_methoden_vergl.eps}
\caption{a) Plot of $F_s(R,\theta)/k_BT$ vs. $R$ for $k_BT/J=3.0$
and a broad range of values for $\theta$. Dotted lines are results
for $L=D=30$, dashes lines for $D=L=40$, dash-dotted lines for
$L=60$, $D=20$. For the systems (bulk, $H_1/J=0$ and $H_1/J=0.4$
$(\theta\approx 58^o)$) the classical predictions $4 \pi R^2
\gamma_{v \ell}$ and $4 \pi R^2 \gamma_{v \ell} f(\theta)$ are
included (full curves).
b) Plot of $\tau$ vs. $\theta$ for
$k_BT/J=3.0$. Crosses are data for $D=L=30$, circles for $D=L=40$,
diamonds $D=10$, $L=60$, and squares for $D=10$, $R=80$.
}
\label{fig4}
\end{figure}

Fig.~\ref{fig4}~a highlights the main results of our study.
For the bulk we can directly test the classical theory
of homogeneous nucleation by comparing our simulation results for
$F_s(R)$ with the capillary approximation $F_{c,bulk}(R)= 4 \pi R^2 
\gamma_{v \ell}$  \cite{11,25} ($\gamma_{v \ell}$ being the interfacial tension for a 
flat interface.) For $R=5$ the difference between the theoretical estimate 
and our simulation results is already less than $5\%$ (at $k_BT/J=3.0$) and quickly
vanishes with increasing droplet radius. The validity of the capillarity approximation 
for the bulk agrees with conclusions drawn from surface force measurements on liquid 
bridges \cite{29a}, but contrasts a large body of other work (see \cite{25} for a discussion).

We have also determined the surface free energy for a sphere-cap shaped 
droplet attached to the wall for several contact angles as a function of the radius of curvature.
It
is important to note that within statistical errors different choices of $L$ and $D$ yield
identical results.
For $\theta=90^o$ and $\theta=58.2^o$ we have also included the theoretical
estimate from the capillarity approximation $F_{c,het}=F_{c,bulk}\cdot f(\theta)$.
When we consider 
$F_{c,het}(R,\theta)-F_s (R,\theta)$, we find that the difference
increases linearly with $R$, and hence can be interpreted as a
line tension contribution, $2 \pi r \tau$. I.e., Eq~\eqref{Turnbull1} is 
modified to $F_{s,het} = F_{s,hom} \cdot f(\theta)+ 2\pi r \tau$. 
This division of surface and line contributions relies on our droplet definition,
Eq~\eqref{eq7}, which uses the fact that for the lattice gas the Tolman length \cite{3,4,5} 
is zero. Fig.~\ref{fig4}b shows a plot of the line tension $\tau$ obtained in this way versus 
contact angle. The line tension is negative, and
becomes very small (and presumably vanishes) as $\theta
\rightarrow 0$. The estimates for $\tau$ have been used (together
with the appropriate estimates for $R$) in Fig.~\ref{fig2}, to
show the expected deviation of the contact angle for
small droplets from the bulk value.
\begin{figure}[t!]
\includegraphics[width=0.8\linewidth]{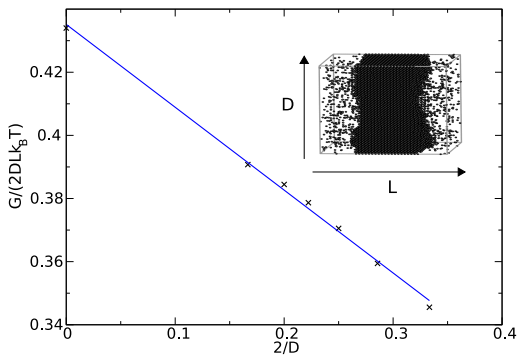}
\caption{Plot of the thermodynamic potential $G$ for slab configurations (insert),
normalized by temperature and total surface area $2LD$, versus $2/D$ so
that the slope yields the line tension ($\tau/k_BT=0.26\pm0.001$). The
intercept of the ordinate was fixed to the literature value of the interface tension 0.434 \cite{17}.
}
\label{fig5}
\end{figure}
An independent check of the line tension is obtained for $\theta = 90^\circ$ by investigating
slab configurations (Fig.\ref{fig5}) with varying D. In this case
the potential can be written as 
$G/(2DL k_B T)=\gamma_{lv} + 2\tau/D$, and the slope in Fig.\ref{fig5} 
agrees with the estimate shown in Fig.\ref{fig4}b. 
Hence, Eqs.~\eqref{Turnbull1} and \eqref{Turnbull2} hold if a correction due to 
line tension is considered. Interestingly, an experimental hint for the need of 
a negative line tension describing bubble formation in water extrusion from mesopores 
was given in \cite{30}.

In summary, we have shown that the classical theory of
heterogeneous nucleation at flat walls, which predicts a reduction
of the free energy barrier by a factor $f(\theta)$, can describe
the actual surface free energies $F_s(R,\theta)$, provided a line tension term is
included into the latter. While the importance
of line tension effects on small nuclei has been stressed in
various other contexts, e.g. \cite{26,27,28}, the present paper
is the first approach which provides a systematic method to obtain both
$F_s(R)$ for spherical droplets in the bulk and $F_s(R,\theta)$
for wall-attached droplets, as well as for the contact angle
$\theta$ and line tension $\tau$. Thus, for the first time we verify 
Turnbull's equation from 1950, 
which still stands as one of the cornerstones 
of the theory of heterogeneous nucleation. 
The methods, which in this letter are described for the simple Ising model, can
readily be generalized to various models in broad classes of 
systems. They will enable significant progress in the understanding
of nucleation phenomena in diverse branches of physics.


\begin{thebibliography}{0}



\bibitem{1} M. Volmer and A. Weber, Z. Chem. Phys. {\bf 119}, 277
(1926)

\bibitem{2} A.C. Zettlemoyer (ed) {\it Nucleation} (M. Dekker, New
York, 1969)

\bibitem{3} K. Binder and D. Stauffer, Adv. Phys. {\bf25}, 343
(1976)

\bibitem{4} P. Debenedetti: {\it Metastable Liquids} Princeton
University Press, Princeton (1997)

\bibitem{5} D. Kashchiev: Nucleation: {\it Basic Theory with
Applications} (Butterworth-Heinemann, Oxford, 2000)

\bibitem{9} J. Curtius, Compt. Rendus Phys. {\bf7}, 1027 (2006)

\bibitem{10} H. Biloni, in {\it Physical Metallurgy} (R.W. Cahn
and P. Haasen, eds.) North-Holland, Amsterdam (1983), p. 477

\bibitem{12} P.R. Ten Wolde and D. Frenkel, J. Chem. Phys.
{\bf109}, 9901 (1998)

\bibitem{27} S. Auer and D. Frenkel, Phys. Rev. Lett. {\bf91},
015703 (2003)

\bibitem{29} L. Maibaum, Phys. Rev. Lett. {\bf 101}, 256102 (2008)

\bibitem{13} T. Young, Phil. Trans. Roy. Soc. London {\bf95}, 65
(1805)

\bibitem{14} P.G.de Gennes, F. Brochard-Wyart, and D. Qu\'er\'e,
{\it Capillary and Wetting Phenomena: Drops, Bubbles, Pearls,
Waves} (Springer, Berlin 2004)

\bibitem{7} R.D. Gretz, J. Chem. Phys. {\bf 45}, 3160 (1966)

\bibitem{6} D. Turnbull, J. Chem. Phys. {\bf18}, 198 (1950); J.
Appl. Phys. {\bf21}, 1022 (1950)

\bibitem{8} G. Navascues and P. Taranzona, J. Chem. Phys. {\bf75},
2441 (1981)

\bibitem{11} D. Winter, P. Virnau,
and K. Binder (in preparation)

\bibitem{21} D. Winter, {\it Diplomarbeit} (Johannes
Gutenberg-Universit\"at Mainz, 2009, unpublished)

\bibitem{18} K.K. Mon, S. Wansleben, D.P. Landau, and K. Binder,
Phys. Rev. B{\bf39}, 7089 (1989)

\bibitem{15} K. Binder and D. P. Landau, Phys. Rev. B{\bf37}, 1745
(1988); K. Binder, D.P. Landau and S. Wansleben, Phys. Rev.
B{\bf40}, 6971 (1989)

\bibitem{16} K. Binder and P.C. Hohenberg, Phys. Rev. B{\bf6},
3461 (1972)

\bibitem{17} M. Hasenbusch, and K. Pinn, Physica A{\bf192}, 342
(1993); {\it ibid} {\bf203}, 189 (1994)

\bibitem{19} B. Widom, J. Chem. Phys. {\bf39}, 2808 (1963)

\bibitem{22} M. Biskup, L. Chayes and R. Kotecky, Europhys. Lett.
{\bf60}, 21 (2002)

\bibitem{23} K. Binder, Physica A{\bf319}, 99 (2003)

\bibitem{24} L.G. MacDowell, P. Virnau, M, M\"uller, and K.
Binder, J. Chem. Phys. {\bf120}, 5293 (2004)

\bibitem{25} M. Schrader, P. Virnau, and K. Binder, Physical Review E {\bf 79}, 061104 (2009) 

\bibitem{26} A. Milchev, and K. Binder, J. Chem. Phys. {\bf114},
8610 (2001)

\bibitem{28} S. Mechkov, G. Oshanin, M. Rauscher, M. Brinkmann,
A.M. Cazabtat and S. Dietrich, Europhys. Lett. {\bf80}, 66002 (2007)

\bibitem{29a} J.Crassous, E.Charlaix, J.L.Loubet, Europhys.Lett. {\bf 28}, 37 (1994)

\bibitem{30} B. Lefevre, A.Saugey, J.L.Barrat, et al., J.Chem.Phys {\bf 120}, 4927 (2004)

\end{thebibliography}
\end{document}